# L'analyse de l'expertise du point de vue de l'ergonomie cognitive

Willemien Visser
EIFFEL2 (Cognition & Coopération en Conception)
INRIA (Institut National de Recherche en Informatique et en Automatique)
Willemien.Visser@inria.fr
http://www-c.inria.fr/eiffel/documents/VISSER%20Willemien%20FICHE%20en.html

**Mots-clés**

Expertise, Recueil et analyse de données, Activité, Structures de représentation et de connaissance, Conception

**Résumé**

Ce texte présente une revue de méthodes pour recueillir et analyser des données sur des actvités complexes. A partir de méthodes développées pour des actvités de conception, nous examinons la possibilité de les transposer à d'autres actvités complexes, notamment des actvités faisant à appel à des expertises sensorielles.

**Abstract**

This paper presents a review of methods for collecting and analysing data on complex activities. Starting with methods developed for design, we examine the possibility to transpose them to other complex activities, especially activities referring to sensorial expertise.

## 1. Introduction

Dans ce texte, nous nous centrons sur des questions méthodologiques. Elles seront illustrées à partir des méthodes développées dans le domaine de l'ergonomie cognitive de la conception (Visser, Darses, & Détienne, 2004), ou, avec une dénomination anglaise plus large, "cognitive design studies" (Visser, 2006a, 2006b). Dans la dernière section du texte, nous discuterons des possibilités de transfert à des expertises sensorielles.

Les premières études sur l'expertise ont porté sur le jeu de grands maîtres joueurs d'échecs (Chase & Simon, 1973; De Groot, 1978). Elles ont utilisé la méthode des protocoles verbaux[1] ou du

---

[1] Dans le cas d'une personne qui réalise seule une tâche que l'analyste cherche à analyser, il s'agit des énoncés que cette personne produit quand elle y a été invitée par le biais d'une consigne qui lui demande de "penser à haute voix" (c'est-à-dire, de procéder à de la "verbalisation") pendant la réalisation de sa tâche (verbalisation 'simultanée'). Dans le cas de personnes travaillant ensemble, il s'agit des énoncés qu'elles produisent spontanément (cette extension du

rappel immédiat[2] pour effectuer des analyses des spécificités cognitives de l'expertise de ces grands maîtres (v. le célèbre article de synthèse sur 40 ans de recherche dans le domaine par Glaser & Chi, 1988). Depuis ces jours-là, de nombreuses approches de l'expertise ont été développées (Bédard & Chi, 1992; Chi, Glaser, & Farr, 1988; Ericsson, 2005, 1996; Green & Gilhooly, 1992; Reimann & Chi, 1989; Rikers & Paas, 2005). Ce texte présente des méthodes de recueil utilisées en psychologie et ergonomie cognitives. Pour des raisons présentées par la suite, les méthodes qui recevront le plus d'attention concernent l'exercice de l'activité qui met en œuvre l'expertise-cible. Pour des présentations plus exhaustives, voir Bisseret, Sebillotte et Falzon (1999 ; v. aussi Kidd, 1987)[3].

*1.1 "Expertise"*

Il y a au moins deux points de vue sur l'expertise. Certains auteurs utilisent le concept pour des performances exceptionnelles dans un domaine—ils qualifient d'"experts", par exemple, seulement des grands maîtres joueurs d'échecs ou des sportifs de haut niveau. Les experts sont des talentueux. Instruction et pratique sont nécessaires, mais pas suffisantes (pour une présentation critique, v. Ericsson & Lehmann, 1996). D'autres chercheurs utilisent le concept de façon plus large. Ils considèrent qu'une personne exerce son expertise dans un domaine, quand, confrontée à une situation, elle dispose des capacités (souvent traduites en termes de "règles et procédures") permettant d'atteindre son objectif, sans, pour ce faire, exploiter des automatismes. Ainsi, "tout un chacun est expert en bien des activités qu'elles soient professionnelles, culturelles [ou] domestiques" (Bisseret et al., 1999, p. 6). Pour Leplat (1986/1992, p. 28), l'"expertise" est un "autre nom de l'habileté quand l'activité s'exerce sur des tâches complexes" (l'"habileté" est aussi souvent appelée "compétence", Angl. *skill*) "

Nous utilisons ici le terme "expertise" dans sa seconde acception. Notre caractérisation assez globale de l'expertise rejoint de nombreuses définitions qui qualifient l'expertise en termes de connaissances et de traitements cognitifs. Elle met toutefois l'accent sur un aspect particulier qui nous paraît spécialement pertinent : "l'expertise dans un domaine réside notamment dans la façon dont un expert *utilise* ses connaissances" (Visser & Falzon, 1988, p. 134). C'est pourquoi dans le reste de ce texte, le terme "expertise" utilisé sans autre précision renverra à la *mise en œuvre* d'une expertise.

Beaucoup d'études concernent l'expertise dans des activités de résolution de problèmes. La majorité sont des activités de jeu, plus ou moins artificielles, ou des tâches construites pour l'étude, mais on a aussi analysé des activités professionnelles, telles que la conception ou le diagnostic médical.

Nous reformulons ici de façon plus générique la caractérisation générale de l'expertise formulée essentiellement sur la base d'études sur la résolution de problèmes (Chi et al., 1988; Glaser & Chi, 1988). Les différences entre experts et novices dans un domaine sont généralement considérées résulter d'une interaction entre l'organisation des connaissances des personnes et leurs capacités particulières de traitement. Les connaissances de l'expert sont organisées sur la base de traits abstraits, conceptuels des entités du domaine, en termes de principes sous-jacents, pertinents pour réaliser la tâche-cible. Des novices s'appuient plutôt sur des traits de surface, des

---

terme "protocole verbal" n'est pas admise par tous les auteurs). Selon les auteurs, le terme "protocole" renvoie aux énoncés ou à leur enregistrement.
[2] Des personnes ("sujets expérimentaux") sont exposées (brièvement) à du "matériel expérimental" (ici des positions d'échecs), après quoi elles sont sollicitées de rappeler ce matériel de façon aussi précise et détaillée que possible.
[3] Pour des méthodes utilisées pour l'étude de l'expertise en conception—ou même spécifiques à ce domaine—, voir Expertise_in_Design_[Special_issue] (2004), Cross (2004b), Visser et Morais (1988) et Visser et Falzon (1988).

critères non spécifiques au domaine et, par conséquent, pas nécessairement pertinents pour la tâche. Les experts sont capables d'inférer des informations absentes des spécifications de sa tâche et, lors de leur analyse de celles-ci, d'en construire une représentation organisée de façon appropriée pour effectuer la tâche. Ils commencent celle-ci d'ailleurs par une analyse qualitative des spécifications, tandis qu'une personne novice a tendance à sauter cette première étape et à s'attaquer directement à l'exécution de la tâche.

La majorité des chercheurs adhèrent à cette caractérisation. Il y a cependant différentes opinions quant à l'origine de l'expertise d'une personne. Renvoyant aux travaux de Galton (1869/1979, cité, par ex., dans Ericsson, 2005), certains auteurs défendent qu'elle dépend de facteurs innés et que l'expérience ou l'apprentissage n'y peuvent rien. Plus récemment, toujours renvoyant aux travaux initiaux de De Groot (1978) et Chase et Simon (1973), les chercheurs estiment plutôt que l'expérience est le facteur décisif. Nous présenterons dans la Discussion brièvement un modèle récent, qui considère une "pratique délibérée" comme élément décisif du développement de l'expertise.

On conçoit généralement que l'expertise est spécifique à un domaine particulier (Chase & Simon, 1973; Ericsson & Lehmann, 1996). Elle n'est donc pas transférable—même pas entre des domaines proches, comme, par exemple, les mathématiques et la physique—sauf en ce qui concerne des éléments partagés.

*1.2 Niveaux d'expertise et Types d'expertise*

Les études classiques sur l'expertise ont souvent procédé à des comparaisons entre experts et novices dans un domaine. La caractérisation générale de l'expertise présentée ci-dessus est basée sur ce type d'études. Elles portent donc sur des *niveaux* d'expertise (Chi et al., 1988; Cross, 2004a; Expertise in Design, 2004; Glaser, 1986; Glaser & Chi, 1988; Reimann & Chi, 1989). D'autre part, il y a des d'études cliniques d'experts, qui permettent d'identifier des caractéristiques d'experts spécifiques (Cross, 2001, 2002).

Nous avons proposé de distinguer de plus des *types* d'expertise (Visser & Falzon, 1988, 1992/3). Des experts avec une longue expérience dans un même domaine, mais ayant rempli différentes tâches à travers leur carrière, déployent en effet différents types de connaissances, dont l'organisation diffère également selon les experts. Notons que, à côté de ces études ayant identifié le rôle de toute une carrière professionnelle, d'autres études ont montré, souvent à travers des expériences de catégorisation, l'influence de tâches spécifiques, limitées dans le temps (Dubois, Bourgine, & Resche-Rigon, 1992/3; Dubois, Fleury, & Mazet, 1993). Un exemple en est l'étude de Gaillard, Billières et Magnen (2005), qui l'a plaidé pour l'"expertise phonétique" dans la discrimination de sons : les performances dans une "tâche [expérimentale] auditive simple de discrimination 'pure'" diffèrent de celles dont on sait qu'elles sont déployées d'habitude dans des activités de tous les jours participant à la communication. D'autres chercheurs, en utilisant des méthodes de classification de "sources odorantes" fondée sur les "airs de famille" de celles-ci, constatent que différentes personnes procèdent à des regroupements différents des odorants (Sicard, Chastrette, & Godinot, 1997; v. aussi Dubois & Grinevald, 1999, pour la désignation de couleurs).

Les études de l'expertise ont généralement été effectuées sur des tâches et activités spécialisées—qu'elles soient professionnelles ou ludiques. Un exemple d'un autre type d'études sont celles qui portent sur la planification dans des situations quotidiennes (planification de repas, Byrne, 1977, ou d'itinéraires). Chalmé a ainsi comparé des "experts experts" et des "experts novices" dans leur planification de trajets en voiture (cf. aussi Hayes-Roth & Hayes-Roth, 1979). Dans ses études, tous les participants étaient en effet "experts" en conduite automobile, mais la moitié était "experts" également dans la connaissance de la région à traverser, tandis que l'autre moitié des participants ne connaissait pas celle-ci (Chalmé, Visser, & Denis, 2004).

Dans le domaine des expertises sensorielles, il semble que les études portent plus souvent sur des expertises mises en œuvre dans des activités "ordinaires", quotidiennes ("de sens commun"; cf. Gaillard et al., 2005)—même s'il y a des exceptions, comme des études sur l'expertise visuelle de dépanneurs (par ex. de circuits electriques), l'expertise olfactive de parfumeurs ou d'"experts en vins".

### 1.3 Activité et Structures de représentation et de connaissance

Il est difficile de caractériser l'expertise (sous-entendue : toute -) qu'une personne a acquise, en général à travers de nombreuses années, dans son domaine d'expertise. Il faut donc faire un choix quant aux aspects sur lesquels on va se focaliser dans ses études de l'expertise—ou avoir une hypothèse forte au sujet de ce qui "fait" l'expertise en question.

Parmi les distinctions possibles, nous privilégions celle entre les structures de représentation et de connaissance ("structures cognitives" dans ce qui suit) et l'activité (la pratique d'une personne) dans laquelle ces structures sont mises en œuvre (Visser, 2006a, 2006b). Considérant que c'est davantage son activité qui caractérise l'expertise d'une personne que les connaissances qu'elle possède (cf. notre définition citée ci-dessus ; cf. aussi les discussions sur la différence entre apprentissage par enseignement scolaire ou académique et apprentissage en tant qu'apprenti dans la pratique, v. Leinhardt, Young, & Merriman, 1995; Schön, 1987), nous adoptons un point de vue dynamique. Cette focalisation sur l'analyse de l'activité a des conséquences pour le choix des méthodes de recueil.

## 2. Méthodes de recueil et d'analyse de données sur l'expertise

Notre point de départ est que l'"utilisation concurrente de différentes méthodes de recueil de données" (Visser & Morais, 1988) est profitable pour l'étude d'activités complexes telles que la conception—et, nous faisons l'hypothèse, de nombreuses activités mettant en œuvre des expertises sensorielles (comme le montrent, par exemple, Van Gog, Paas, & Van Merriënboer, 2005, pour l'expertise en dépannage de circuits electriques examinée à partir de verbalisations et de mouvements oculaires ; v. aussi Ericsson, 2005). L'un des principaux facteurs qui guident alors le choix des méthodes est l'objectif poursuivi.

### 2.1 Recueil en situation et hors situation

Pour ce qui concerne la mise en œuvre de l"expertise, les études menées en situation ("sur le terrain") sont en règle générale plus riches que celles conduites hors situation ("hors contexte"). Hors situation, on peut "se faire une idée" des activités d'une personne et recueillir des données sur les structures cognitives qu'elle *possède*, mais on n'aura pas accès aux modalités de leur mise en œuvre[4].

A travers nos études sur la conception, nous avons conclu que, pour le recueil de données sur la mise en œuvre d'une expertise dans des activités complexes comme celles de conception, les études menées en situation ne sont pas seulement plus riches, mais clairement plus appropriées.

---

[4] Par cette remarque, nous ne voulons pas déprécier les nombreuses études sur les expertises sensorielles conduites hors situation. Elles analysent l'expertise sous d'autres angles que sa mise en œuvre en situation—qui est l'aspect que nous privilégions. Il se peut aussi que les entretiens d'explicitation permettent d'accéder à des données sur l'activité (v. la suite du texte).

Nous avons observé en effet que certains aspects esentiels de l'expertise en conception sont déformés hors situation. Il en va ainsi de la façon dont les concepteurs organisent leur activité. Hors situation, les concepteurs présentent leur activité comme étant bien structurée, tandis que, en situation, on observe que l'activité effective a un caractère "opportuniste" (Bisseret, Figeac-Letang, & Falzon, 1988; Guindon, 1990; Visser, 1994b). "Même" des concepteurs professionnels experts, travaillant dans leur environnement de travail habituel, ne suivent pas systématiquement les plan préétablis qu'ils possèdent en mémoire. Au contraire, ils exploitent certaines "opportunités", c'est-à-dire des occasions de traitement d'information qu'ils identifient comme potentiellement intéressantes d'un point de vue de leur "coût cognitif" (Visser, 1994a).
Notre accent sur l'activité mérite au moins deux remarques. Premièrement, il est évident que les activités et les structures cognitives ne se distinguent que de façon analytique : les activités "utilisent" les structures, qui se construisent à travers des activités. Deuxièmement, en examinant des travaux qui présentent des caractérisations de structures cognitives (expertes ou non), il est utile de se demander si (ces travaux permettent de savoir si) ces structures sont *effectivement utilisées* dans l'activité ou si (tout ce que ces travaux permettent de savoir est que) il s'agit de structures qu'une personne experte *possède*.

   *2.2 Objectifs du recueil*

Quant au choix des méthodes de recueil, nous ne distinguons ici que trois objectifs globaux : une première familiarisation avec le domaine d'expertise, le recueil de données sur les connaissances que possède une personne experte (qu'elle les utilise ou non) et le recueil de données sur l'activité d'une personne experte (c'est-à-dire, la mise en œuvre de son expertise). Après quelques observations concernant les deux premiers objectifs, les méthodes servant le dernier objectif recevront le plus d'attention.

**2.2.1 Première familiarisation avec le domaine d'expertise**
*Analyse de la tâche*. En ergonomie, avant de procéder à l'analyse de l'activité d'une personne (experte ou non), on fait généralement une analyse de la tâche de la personne[5]. Cette analyse, qui peut être plus ou moins globale[6], permet aux chercheurs de se familiariser avec les spécificités d'un domaine qu'ils ne connaissent peu ou pas en général. Par ailleurs, elle peut fournir des données sur les grandes lignes de l'activité, notamment sur les conditions dans lesquelles l'activité s'exercera.
L'analyse de la tâche fait en général appel à des entretiens, qui sont souvent semi-dirigés, à partir de questions pré-établies (par exemple, Visser & Morais, 1988; pour d'autres méthodes, v. Bisseret et al., 1999; Visser & Falzon, 1988; Visser & Morais, 1988). Si l'on peut procéder à des observations en situation, ces entretiens sont effectués en général avant d'aller effectuer des observations sur le terrain. Après, les chercheurs ressentent souvent le besoin d'éclaircissements et retournent voir les experts—si possible.

---

[5] Par "tâche", l'ergonome renvoie au but à atteindre par une personne et les conditions dans lesquelles ce but doit être atteint : la tâche peut être *prescrite* (par la hiérarchie et/ou les concepteurs du système de travail) ou la tâche *effective*, celle que la personne se donne elle-même. Par "activité", l'ergonome renvoie à ce que la personne met en œuvre pour exécuter la tâche qu'elle s'est fixée (Leplat, 1986/1992; Leplat & Hoc, 1983). Quand ceci concerne le niveau cognitif, on parle d'"activité cognitive". Il s'agit de la façon dont une personne utilise ses connaissances et autres sources d'information auxquelles elle fait appel. L'activité diffère toujours de la tâche, même de la tâche effective ; l'ergonome examine souvent ces différences.
[6] Toutes les analyses d'une activité peuvent avoir, par ailleurs, un grain plus ou moins fin (cf. l'opposition entre approches macroscopique et microscopique, Garrigou & Visser, 1998).

En ce qui concerne les connaissances, les entretiens traditionnels en ergonomie ne permettent de recueillir que des données sur des connaissances que les experts possèdent, mais n'utilisent pas nécessairement. Ils ne sont donc pas appropriés pour recueillir des données sur les activités cognitives, les connaissances mises en œuvre effectivement et leur mise en œuvre. Ces informations sur des aspects dynamiques de l'expertise doivent être inférées à partir de l'analyse de l'activité.

Depuis une dizaine d'années, Vermersch (1994) développe une nouvelle méthode d'entretien, l'"entretien d'explicitation". L'une des hypothèses de ces entretiens, qui sont bien conduits plus ou moins de temps après la situation visée, est qu'ils peuvent donner accès aux structures (cognitives, mais aussi émotives) qui étaient actives *dans* la situation. C'est que l'entretien d'explicitation commence par une remise "en situation" de la personne interrogée. Cette méthode n'est pas encore très courante en psychologie ergonomique.

Dans le domaine sensoriel (olfactif, acoustique, visuel), de nombreuses données ont été recueillies par le biais de questionnaires, sur différentes formes linguistiques évoquées par des (représentations de) phénomènes sensoriels. A partir des différents types d'expressions linguistiques identifiées, les chercheurs formulent des hypothèses sur l'expérience que les personnes interrogées font de ces phénomènes (Dubois, 2000; Intellectica, 1997; Valentin, Chollet, & Abdi, décembre 2003).

Des analyses globales de la tâche peuvent aussi servir à sélectionner une ou plusieurs parties de la tâche ou phases de l'activité pour les analyses approfondies des aspects de l'expertise sur lesquels on veut se focaliser[7].

### 2.2.2 Recueil de données sur les connaissances qu'une personne experte possède

Il y a de nombreuses méthodes pour recueillir des données sur les connaissances qu'une personne experte possède. Nous présentons deux exemples utilisées dans des études sur la conception (pour des présentations plus exhaustives, v. Bisseret et al., 1999; pour des présentations de méthodes utilisées dans le domaine des expertises sensorielles—surtout sur les expressions langagières—v. Dubois, 2000; Intellectica, 1997).

*Analyses du produit final d'une activité*. Les chercheurs peuvent analyser ce produit (par ex., le logiciel conçu par des concepteurs de logiciel, Visser & Morais, 1988) ou en comparer deux versions, l'une étant conçue par les concepteurs sur la base des spécifications reçues, l'autre par les chercheurs sur la base des règles inférées de celles-ci (Visser, 1985). Les concepteurs peuvent aussi l'analyser eux-mêmes (commentaires de programmes dans Visser, 1985).

*Etudes expérimentales de catégorisations effectuées par différents experts*. Sur la base des catégorisations établies par différents experts et des différences constatées entr'elles, on peut formuler des hypothèses sur les activités qui ont conduit à ces résultats. Utilisant une consigne "mettant en situation" les experts, on peut ensuite formuler également des hypothèses au sujet de l'influence des conditions dans lesquelles la catégorisation s'est effectuée (les différentes tâches et activités des experts) sur l'organisation et la nature des catégorisations. Plusieurs études ont ainsi conclu au rôle décisif des objectifs que les experts poursuivent dans leurs tâches (Visser & Falzon, 1992/3; Visser & Perron, 1996). Ce type d'études sont aussi très fréquentes pour l'analyse des expertises sensorielles (Dubois & Grinevald, 1999).

---

[7] Nous avons proposé de réserver "étapes" pour des périodes de la tâche prescrites par les méthdologies utilisées (des sous-tâches), et "phases" pour des périodes dans l'activité qui constituent des entités disjointes d'un point de vue cognitif.

L'observation d'experts en situation n'est pas toujours possible. Un exemple en est la situation de "sauvetage" de l'expertise d'une personne partant à la retraite qui n'est plus chargée de nouveaux projets. Il est impossible alors de suivre la personne pendant son activité experte. Nous avons développé un certain nombre de méthodes spécifiques pour cette situation, pour recueillir des données sur les raisonnements utilisés par l'expert et les concepts utilisés dans ces raisonnements (Visser & Falzon, 1988, 1992/3).

**2.2.3 Recueil et analyse de données sur la mise en œuvre d'une expertise**
A travers nos études sur la conception, nous avons donc conclu que pour des activités complexes comme celles de conception, la meilleure situation pour le recueil de données sur l'expertise est une situation dans laquelle la personne exerce cette activité. On peut alors observer l'expert dans toute la richesse de son activité. L'artifice majeur dans ses études observationnelles est souvent l'appel que les chercheurs font à la verbalisation.

*Verbalisation : protocole verbal et enregistrement vidéo*. L'observation d'une personne experte en situation est une méthode lourde et coûteuse. Classiquement on recueille un protocole verbal auprès de cette personne (trace de ses verbalisations simultanées), à l'époque actuelle souvent accompagnée de l'enregistrement vidéo de la situation observée. La verbalisation peut être concomitante ou différée (consécutive ou rétrospective, assistée en général par des traces de l'activité). L'analyse d'un tel protocole prend beaucoup de temps (le texte de référence dans ce domaine est Ericsson & Simon, 1993)
Si l'on étudie une personne qui effectue son activité seule (par exemple, travaillant sur un projet de conception mené individuellement), il faut demander explicitement à la personne de verbaliser (c'est-à-dire, d'énoncer, à haute voix, toutes les pensées qui lui traversent la tête ; la verbalisation est "provoquée"), tandis que des personnes travaillant ensemble procèderont en général à des polylogues, sans qu'on ait à les y inviter (pour une méthode d'analyse de ce type de données, v. Darses, Détienne, Falzon, & Visser, 2001).

*Prise de notes et recueil des traces.* La situation observée peut avoir une telle durée qu'il est irréaliste d'enregistrer "tout". La personne qui observe peut alors procéder à la prise de notes et le recueil des traces de l'activité (pour des exemples du type de note et traces, v. Visser, 1994a; Visser, 1994b). Toute observation se fait de façon armée : la personne qui effectue une étude a un cadre théorique et un objectif qui la guident dans ses choix d'observations—et leur interprétation. Ce caractère armé de l'observation est très prégnant dans une situation où le recueil s'effectue de façon sélective par prise de notes, mais on fait aussi ses choix quand on enregistre "tout".

*Situation expérimentale contrôlée*. Même si nous avons insisté beaucoup sur le besoin d'observations effectuées sur le terrain (c'est-à-dire, pour la conception, dans des projets professionnels industriels), le recueil de données pertinentes peut se faire dans des situations expérimentales contrôlées "écologiques" (par exemple, la "Delft study", dans laquelle un groupe de chercheurs en "cognitive design studies" a pu analyser les données provenant d'un projet de conception réalisé aussi bien par des concepteurs travaillant individuellement que par des concepteurs en équipe, v. Cross, Christiaans, & Dorst, 1996).

*Analyse des données*. L'analyse des données et bien sûr au moins aussi important que le recueil. Pourtant, nous ne pouvons le détailler ici, vu l'espace dont nous disposons. On utilise des méthodes quantitatives et qualitatives, souvent sur les mêmes corpus.

**3. Discussion**

Nous avons insisté dans ce texte surtout sur les différentes possibilités pour recueillir des données sur des expertises telles qu'elles sont mises en œuvre *effectivement* dans des activités complexes, partant de l'idée qu'elles peuvent aussi être profitables pour l'étude d'expertises sensorielles dans des activités d'une autre nature. L'approche privilégiée ici, à savoir "le terrain", nous paraît en effet s'imposer pour ces études—en tout cas si elles visent l'accès à l'expertise d'un point de vue de *l'expérience de la personne experte*, plutôt que du point de vue des conditions physiques "objectives". Ce point de départ demande toutefois une discussion. Nous y apportons ici deux commentaires. Le dernier point de discussion concerne une autre approche de l'étude de l'expertise.

*3.1 Modalités d'observation pour l'étude de l'expertise*

L'utilisation de la méthode des protocoles verbaux—qui accompagne souvent les études observationnelles—est soumise à des conditions. Au moins deux sont particulièrement critiques. Les données traitées par la personne à qui l'on demande de "penser à haute voix" doivent être "directement disponibles sous une forme propositionelle", c'est-à-dire, langagière ou verbale (Ericsson & Simon, 1993, p. 235). D'autre part, "seulement l'information sur laquelle l'attention est focalisée peut être exprimée" dans la verbalisation (ibid.). Ceci rend la méthode donc inadaptée à l'étude d'activités automatisées et lui enlève beaucoup d'intérêt pour l'étude d'expertises sensorielles—au moins pour des aspects importants d'elles, dont on peut supposer que les processus correspondants ne s'appuient pas sur des données ayant une "forme propositionelle".
Le choix d'une approche dynamique est étayée toutefois par des études qui montrent que les performances dans certaines tâches (expérimentales) dans le domaine d'expertise sont de meilleurs prédicteurs de la performance experte que la performance mnésique (mise en œuvre dans le rappel immédiat ou la catégorisation) (Ericsson, 2005). Ericsson propose que l'on identifie des tâches qui captent l'essence de l'expertise dans un domaine et que l'on analyse (par ex. en utilisant des méthodes de traçage de processus, Angl. *process-tracing methods*) les mécanismes qui interviennent dans la performance supérieure des experts dans ce domaine. Notons que l'analyse de verbalisations simultanées ou consécutives ne donne pas accès aux mêmes structures que celle des expressions langagières dans des tâches de catégorisation ou autres tâches évoquant ou provoquant ces expressions—l'une des approches le plus fréquentes dans les études des expertises sensorielles (v., par ex., la comparaison entre des "experts en vin" et des novices, Ballester et al., 2005; v. aussi Valentin et al., décembre 2003).

*3.2 Prise en compte de composantes non verbales*

Les expertises sensorielles font donc intervenir d'autres modalités que la parole. Ceci vaut également pour de nombreuses d'activités de conception. La conception architecturale (coopérative), par exemple, utilise beaucoup l'expression graphique et gestuelle. Nous avons développé un langage de description pour les modalités correspondantes (Détienne, Visser, & Tabary, à paraître; Visser & Détienne, 2005), ce qui nous a permis d'identifier différents modes d'articulation de ces deux composantes sémiotiques des interactions entre des architectes (Détienne & Visser, 2006). Nous ne sommes pas consciente de méthodes permettant l'identification des modalités d'expression sémiotique des activités mettant en œuvre des expertises sensorielles. On peut penser à des entretiens d'explicitation ou des verbalisations consécutives assistées par des traces de l'activité.

*3.3 Etude de l'expertise en vue d'un entrainement à l'expertise*

Récemment, des études dans le courant de la "pratique délibérée" (Angl. *deliberate practice*)—sous-entendu, d'aspects particuliers de l'expertise—commencent à identifier des activités qui correspondent à ces aspects et dont la pratique—de façon "massive"—permet à des personnes qui ne sont pas (encore) expertes d'améliorer leur niveau d'expertise, un aspect à la fois (Ericsson, 2005). Quelques exemples de telles activités sont la sélection des meilleurs mouvements dans des parties d'échecs historiques, la planification mentale et l'auto-explication de processus de résolution de problème. Si le but de recherches sur l'expertise est l'identification de moyens permettant d'amener des personnes à l'expertise, l'identification de telles activités peut être un objet de recherche pertinent.

**Références bibliographiques**